\documentclass[12pt]{article}

\usepackage{multicol}

\usepackage{geometry}
\usepackage{graphicx}
\usepackage{newtxtext}
\usepackage{newtxmath}
\usepackage{hyperref}

\usepackage{amsmath}
\usepackage{cuted}

\usepackage[utf8]{inputenc}

\usepackage{xcolor}
\usepackage[ruled,vlined]{algorithm2e}

\SetCommentSty{mycommfont}
\usepackage{bm}
\usepackage{cases}
\graphicspath{{././}}
\usepackage[section]{placeins}
\usepackage{siunitx}

\hypersetup{
    colorlinks = true,
    urlcolor   = blue,
    citecolor  = black,
}

\newcommand{\RomanNumeralCaps}[1]
\linenumbers

\newcommand{\mob}{M\"obius}

\newcommand{\tresi}{\mathbb{R}}
\newcommand{\tmob}{\mathbb{M}}
\newcommand{\tmobc}{\boldsymbol{\mathcal{C}}}
\newcommand{\tmobr}{\boldsymbol{\mathcal{R}}}
\newcommand{\tmobrc}{{\mathcal{R}}}

\newcommand{\tch}{\boldsymbol{\mathsf{Ch}}}

\newcommand{\tpt}{triply-twisted}

\newcommand{\bsc}{binormal scroll}

\newcommand{\deve}{\text{D}}
\newcommand{\bscr}{\text{B}}
\newcommand{\roto}{\text{R}}

\newcommand{\otwist}{\omega_{\text{twist}}}

\newcommand{\sgn}{\mathop{\mathrm{sgn}}}

\newcommand{\trot}{\boldsymbol{\mathsf{\Omega}}_o}

\newcommand{\vo}{{\bm v}_o}
\newcommand{\bomega}{\boldsymbol{\omega}}
\newcommand{\tauo}{\boldsymbol{\tau}_o}
\newcommand{\bff}{\boldsymbol{f}}
\newcommand{\ttran}{\boldsymbol{\mathsf{K}}}
\newcommand{\tcou}{\boldsymbol{\mathsf{C}}_o}
\newcommand{\tcout}{\boldsymbol{\mathsf{C}}_o^{\scriptstyle\top}}

\newcommand{\trotc}{{\mathsf{\Omega}}_o}
\newcommand{\ttranc}{{\mathsf{K}}}
\newcommand{\tcouc}{{\mathsf{C}}_o}

\newcommand{\bfe}{\boldsymbol{e}}

\newcommand{\Ar}{\textsf{A}\mskip0.625mu\textsf{r}}
\renewcommand{\Re}{\textsf{R}\mskip0.25mu\textsf{e}}
\newcommand{\Rer}{\textsf{R}\mskip0.25mu\textsf{e}_r}

\makeatletter
\newcommand*{\bigcdot}{}
\DeclareRobustCommand*{\bigcdot}{%
  \mathbin{\mathpalette\bigcdot@{}}%
}
\newcommand*{\bigcdot@scalefactor}{.5}
\newcommand*{\bigcdot@widthfactor}{1.15}
\newcommand*{\bigcdot@}[2]{%
  \sbox0{$#1\vcenter{}$}
  \sbox2{$#1\cdot\m@th$}%
  \hbox to \bigcdot@widthfactor\wd2{%
    \hfil
    \raise\ht0\hbox{%
      \scalebox{\bigcdot@scalefactor}{%
        \lower\ht0\hbox{$#1\bullet\m@th$}%
      }%
    }%
    \hfil
  }%
}
\makeatother

\definecolor{newred}{RGB}{180,20,5}

\definecolor{newgreen}{RGB}{1,129,30}

\definecolor{mybluei}{RGB}{28,138,207}
\definecolor{myblueii}{RGB}{131,197,231}
\definecolor{newBlue}{RGB}{10,10,255}

\makeatletter							
\def\printtitle{
	{\centering \Large \sc \textbf{\@title}\par}}		
\makeatother	

\makeatletter							
\def\printauthor{
	\vspace{0.5cm}
 {\centering \small \@author}}				
\makeatother

\title{Sedimentation dynamics of triply-twisted \mob\ bands: Geometry versus topology}

\author{%
Nicolas Moreno$^{1}*$, David Vazquez-Cortes$^{2}$, and  Eliot Fried$^{2}\dagger$ \\
{1.Basque Center for Applied Mathematics (BCAM), Alameda de Mazarredo 14, Bilbao 48400, Spain}\\
{2. Materials and Mechanics Unit, Okinawa Institute of Science and Technology, 1919-1 Tancha,Onna-son, 904-0412, Okinawa, Japan}\\
{$*$ nmoreno@bcamath.org  -- $\dagger$ eliot.fried@oist.jp}\\
}
\date{2024}

\begin{document}

\printtitle 
	
	\printauthor

\begin{abstract}
Chiral objects have intrigued scientists across several disciplines, including mathematics, crystallography, chemistry, and biology. A M\"obius band, an emblematic chiral structure, can be made by connecting the ends of a strip after applying an odd number of twists. Traditionally, the direction of twist governs its rotational behavior during sedimentation in a fluid. Here, we present experimental and computational investigations of triply-twisted M\"obius bands boasting threefold-rotational symmetry that challenge this prevailing understanding. We explore three types of bands with different curvature, each defined by its construction method.  Experimental observations reveal that all three types of bands align axially and exhibit rotational motion during sedimentation. Surprisingly, for only one type of band the spinning direction (chiral hydrodynamic response) departs from expectations; it is not solely determined by the twist direction but changes with the aspect ratio of the band. Numerical simulations corroborate this observation, and in-depth analysis of the resistance tensors of each type of band sheds light on the possible causes of this transition. We propose that modifications in fluid-induced drag, combined with inertial effects, underpin this  phenomenon. Our study challenges existing knowledge of chiral object hydrodynamics, enriching our understanding of complex fluid dynamics. Moreover, it offers transformative potential across diverse fields, promising advancements in mixing, separation processes, and innovative passive swimmers.
\end{abstract}

Molecular chirality is essential in  many scientific and industrial settings, shaping the properties and applications of a vast array of substances. In pharmaceuticals, molecular chirality is pivotal, influencing the effectiveness and safety of drugs.\cite{Burke2002,Nguyen2006} This principle extends to agrochemicals, where enantiomers may exhibit vastly different biological activities, impacting their efficacy in pest control.\cite{Liu2005,Jeschke2018} In the flavor and fragrance industries, chirality is leveraged to craft distinct olfactory experiences, since enantiomers can impart different smells and tastes.\cite{Engel2020,Michailidou2023} Display technologies rely crucially on chiral liquid crystalline molecules.\cite{Tamaoki2001,Bisoyi2022} Chirality is also a consideration in the cosmetics industry, as enantiomers of skincare ingredients can elicit varied effects.\cite{Kandula2023} The identification and separation of enantiomers are crucial in these applications. Whether ensuring the safety and efficacy of pharmaceuticals, optimizing the performance of agrochemicals, crafting flavors or fragrances, or tailoring materials with desired characteristics, the ability to discriminate and isolate enantiomers is foundational to harnessing the full potential of molecular chirality.  Functional materials with tailored chirality have  recently been successfully synthesized.
\cite{Geng2019,Ouyang2020,Zhang2016,Fan2023} These materials have applications to chiral sensing,\cite{Yoo2019} catalysis,\cite{Zhang2020} and nanoscale flow-driven rotary motors.\cite{Shi2022} Various studies have demonstrated the viability of leveraging hydrodynamic effects for the separation of chiral molecules in solution.\cite{Kim1991,Meinhardt2012,Hermans2015,Clemens2015,Doi2016} Our understanding of how chiral objects respond to various external physical and chemical stimuli is still incomplete, presenting a complex challenge.

In a low Reynolds number regime, a rigid body moving in an incompressible, viscous liquid experiences hydrodynamic forces and torques,  leading to translation and rotation. The coupling of these motions depends on the geometry and topology of the body. For instance, a screw-like rigid body rotating in a quiescent fluid imparts linear motion relative to the fluid, generating a force, and simultaneously induces torque, leading to vortical motion in the liquid.\cite{Doi2016,Makino2005,Witten2020b} Similar phenomena are observed in biological structures, where chirality is  linked to form.\cite{Krapf2009} As the Reynolds number increases within the laminar range (say from $10^2$ to $4\times10^2$), wake formation and other hydrodynamic effects can give rise to intricate oscillatory trajectories of the body\cite{Witten2020b}. The stability of these trajectories is intricately linked to the specific geometrical and topological attributes of the considered body.

\begin{figure*}[!t]
\centering
\includegraphics[trim=0cm 0cm 0cm 0cm, clip=true]{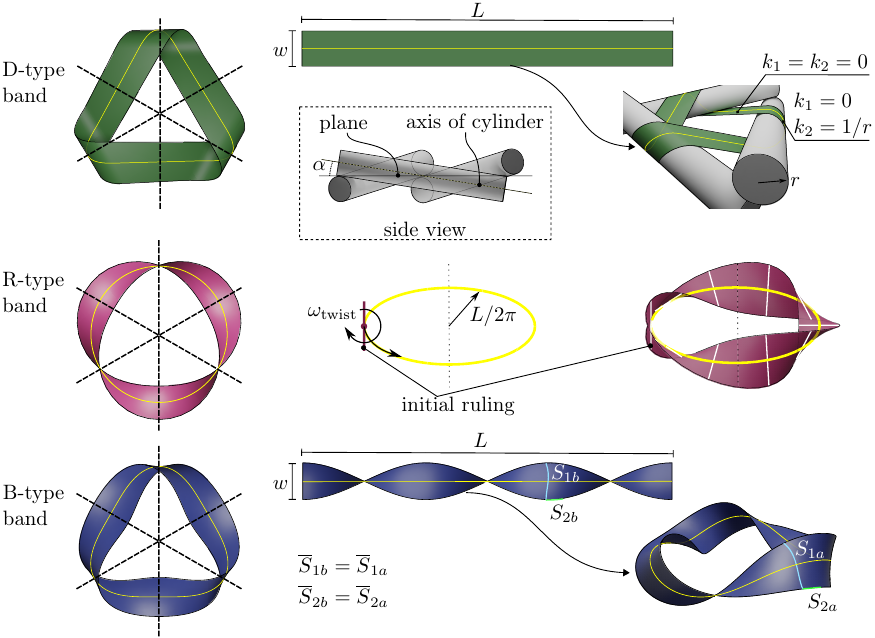}
\caption{ 
Triply twisted, three-fold symmetric, \mob\ bands and their construction. Bands of type developable (\deve) are constructed by bending a rectangular strip of length $L$ and width $w$ around three cylinders of radius $r$, and angle $\alpha$ between them. The principal curvatures $k_1$ and $k_2$ are also shown for the different sections of the band. Bands of type rotoidal (\roto) are constructed by moving a line segment, from its middle point, around a circle of radius $L/2\pi$ while turning it perpendicular to the movement at a rate twist of $\omega_\text{twist}=3\pi/L$. Construction of bands of type binomial-scroll are created by isometrically bending a helicoid, as explained in \cite{Chaurasia2023}. Consequently, the length of curves $S_{1b}$ and $S_{2b}$ on the helicoid remains unchanged after forming the three-twist \mob\ band. The midline of the three types of bands is presented in yellow.
}
	\label{fig:bandsDescrip}
\end{figure*}

Happel and Brenner provided a concise framework for characterizing the dynamics of an arbitrarily shaped rigid body moving in a viscous incompressible liquid.\cite{Happel1981} Given a point $o$ affixed to the body, let $\vo$ be the instantaneous velocity of that point, and let $\bomega$ be the instantaneous angular velocity of the body about that point. The net  force $\bff$ and net hydrodynamic $\tauo$ exerted by the liquid on the body are then given by
\begin{equation}
\begin{bmatrix}
\bff
\\
\tauo
\end{bmatrix}
=-\mu\mskip1.5mu\tresi
\begin{bmatrix}
\vo
\\
\bomega
\end{bmatrix},
\quad
\tresi=
\begin{bmatrix}
\ttran & \tcout
\\
\tcou & \trot
\end{bmatrix},
\end{equation}
where $\mu$ is the viscosity of the liquid and $\ttran$, $\trot$, and $\tcou$ are the translation, rotation, and the coupling dyads. While $\ttran$ is an intrinsic property determined entirely by the geometry and topology of the body, $\trot$ and $\tcou$ depend also on the choice of the origin $o$. The dyads $\ttran$ and $\trot$ are necessarily symmetric and positive-definite. Although $\tcou$ is generally asymmetric, the resistance tensor $\tresi$ is consequently symmetric and positive-definite.  For a chiral body $\tresi$ encodes its preferred spinning direction.\cite{Doi2016,Morozov2017, Mirzae2018} We examine three distinct families of \tpt, three-fold symmetric \mob\ bands, investigating the variability of their unique intrinsic chiral response within each family based on two dimensionless parameters: a notion of aspect ratio that we define subsequently and the Archimedes number $\Ar$ (the ratio of gravitational forces to viscous forces). \mob\ bands can be constructed in various ways. One approach includes bending a rectangular strip of length $L$ and width $w$ into a sequence of three helical sections linked by planar segments,\cite{Schonke2021} as illustrated in Fig. \ref{fig:bandsDescrip}. Any such band is \textit{developable}, meaning that it can be isometrically flattened without stretching, contraction, folding, or tearing. We use \deve\ to denote the family of ruled \tpt\ \mob\ bands constructed in this way. We also consider \textit{rotoidal bands}, constructed by uniformly translating a line segment of length $w$ along a circle with a radius of $L/2\pi$, while simultaneously rotating it at a twist rate of $\omega_{\text{twist}}=\pm3\pi/L$,\cite{Piette2020} as illustrated in Fig. \ref{fig:bandsDescrip}. To ensure the resulting surface remains free from self-intersections, we require that $w<L/\pi$. These bands belong to the family denoted as \roto\ within the category of as illustrated in Fig. \ref{fig:bandsDescrip}, \mob\ bands. We also consider  a class of ruled, \tpt, three-fold symmetric \mob\ bands which arise as the limiting surfaces of closed kinematic chains known as \mob\ kaleidocycles, similar to the binormal-scroll (\bscr) bands\cite{Schonke2019}. These \mob\ bands are isometric deformations of helicoids,\cite{Chaurasia2023} exclusively attainable through bending and twisting, without involving any stretching or contraction. While each \mob\ band considered has the same topology and has three-fold rotational symmetry, each family of \mob\ bands has distinct features which we describe in the next two paragraphs. 
%
%
%

%

The dynamics of a rigid body sedimenting in a liquid are significantly influenced by the angles at which the liquid strikes the body. It is therefore important to consider the projected area and curvature of the sedimenting body as pivotal factors. While maintaining a fixed aspect ratio, the projected area can differ across the three families of bands considered. To enable a meaningful comparison, we introduce the notion of the `projected radius', denoted as $R$, which corresponds to the radius of the circle inscribed within the projected midline of a band. The projected midline takes the form of a triangular loop for \deve\ bands, a circle for \roto\ bands, and what resembles a Reuleaux for \bscr\ bands, as illustrated in Fig. \ref{fig:bandsDescrip} and SI Fig. 3. We consider it noteworthy that for a constant $R$ size, \deve\ bands are more compact than \bscr\ bands with the same aspect ratio, whereas \bscr\ bands are notably more compact than \roto\ counterparts with the same aspect ratio. In a broader sense, \roto\ bands can achieve higher aspect ratios because their circular midlines provide more internal space within the loop. Conversely, \deve\ and \bscr\ bands are expected to have smaller aspect ratios owing to their relatively confined inner available space. Considering the curvature characteristics of the bands, \bscr\ and \deve\ bands differ due to disparities in the rates at which their rulings rotate around their midlines. Specifically, \deve\ bands exhibit zero Gaussian curvature ($K=0$) and comprise three straight segments where rotation rates are null. In contrast, \bscr\ and \roto\ bands share negative Gaussian curvature ($K<0$) attributed to the constant twist of the generatrix, which we denote by $\otwist$ ($8.0941/L$\cite{Schonke2019} and $3\pi/(L)$ for \bscr\ and \roto\ bands respectively). We speculate that \bscr\ bands amalgamate select geometrical properties from \deve\ and \roto\ bands in novel ways, underpinning their behavior during sedimentation. Moreover, these distinctions influence the resistance tensor of \bscr\ bands, leading to intricate chiral transitions in their hydrodynamic response.

\begin{figure}[!t]
	\centering
\includegraphics[trim=0cm 0cm 0cm 0cm, clip=true]{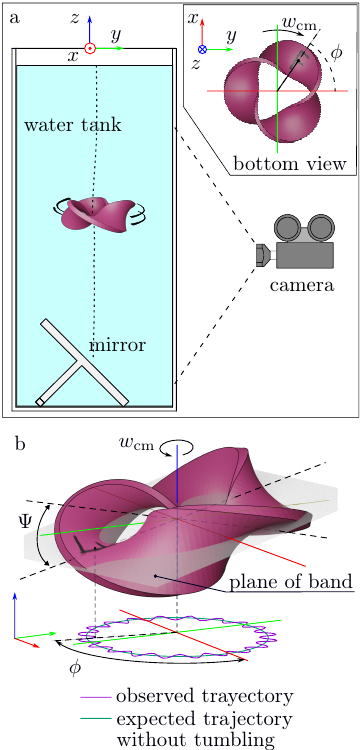}
\caption{
Schematic of experimental setup to study the sedimentation of triply twisted \mob\ bands. a) In the base of a water tank with a square cross-sectional area, a mirror is positioned. The camera in front of the water tank allows us to directly observe the movement of \mob\ bands in the $(x,z)$-plane during sedimentation. The mirror at the bottom of the tank allows us to observe the movements of the bands in the $(x,y)$-plane as sketched in the bottom-view inset. b) The \mob\ band rotates at angular velocity $\omega_\text{cm}=\dot{\phi}$ and tumble at angle $\Psi$. The red, green, and blue axis are parallel to the $x$, $y$, and $z$ axis, respectively, and pass through the center of mass of the band. $\phi$ is the angle formed by the red axis and the projection in the $(x,y)$-plane of the line segment connecting the center of gravity of the band and the position of the marker ``F''
}
\label{fig:systemDef}
\end{figure}

We conducted comprehensive sedimentation studies, encompassing both experimental and numerical investigations, focusing on \tpt\ bands. These three-fold symmetric bands are characterized by their projected radius $R$, width $w$, and mass density $\rho_s$. Sedimentation occurred in an incompressible, viscous liquid with mass density $\rho$ and shear viscosity $\mu$. The width $w$ of the bands varied among different cases to explore the effects of aspect ratio. In Fig. \ref{fig:systemDef}, we provide an illustration of the sedimentation setup and the various motions experienced by the bands during the experiments. Our experimental studies used 3D-printed polystyrene bands, each with $R=1$~cm, sedimenting in water.  See Supplementary Materials for samples of stl files used to print the bands. Further details on the experimental setup can be found in the Methods section. In this context, the forces influencing the bands include the force due to gravity, the buoyancy force, and the drag force. As each band descends, it eventually reaches a terminal velocity $v_z$, measured with respect to a reference frame located at the top of the container. Additionally, they exhibit an angular velocity, $\omega_{\text{cm}}$, relative to a moving frame fixed on the center of mass of each band and aligned with a suitably chosen frame of reference determined by the container. The sign for spin or chiral response is designated as positive ($+$) when the band rotates counter-clockwise, as observed from a top-down perspective. For these bands, we define the geometric or intrinsic chirality based on the direction of twist ($w_\text{twist}$). For further clarification, please refer to SI.Fig. 1, which includes an illustrative example contrasting two different enantiomers derived from the same midline.

To facilitate a comparative study of the influence of aspect ratio $w/R$ across the three types of \mob\ bands, we introduce the parameter $\chi = w/w_{\text{max}}$, with $w_{\text{max}}$ representing the width when the edges touch at the center (details provided in SI.Fig. 2). Bands of type \roto, characterized by higher attainable aspect ratios $w/R$, exhibit the maximum surface area at $\chi_{\text{max}}$. For bands of type \bscr\ and \roto, $\chi$ varies within the range $0.1\le\chi\le1$. In comparison, the construction of bands of type \deve\ involves two additional parameters: the radius $r$ of the three cylinders about which the curved sections of the band are wrapped, and the angle $\alpha$ between these cylinders,\cite{Schonke2021} as illustrated in Fig. \ref{fig:bandsDescrip}. To streamline our investigation, we set $\chi=0.6$ and $\chi=0.8$ for bands of type \deve\ while allowing for variations of the cylinders angle. In presenting our findings, we utilize the following dimensionless quantities: the translational Reynolds number ($\Re={v_z\rho R}/\mu$), the \textit{signed} rotational Reynolds number $\Rer = {\omega_{\text{cm}}\rho R^2}/\mu$, and the Archimedes number $\Ar={\rho(\rho-\rho_s)gR^3}/\mu^2$.

\subsection*{Results and Discussion}

We observed that all three types of bands align axially\cite{Moths2013}, exhibiting a preferred orientation during sedimentation, regardless of their initial orientation. This behavior is consistent with the existence of a stable dominant mode in the translational response of the bands.\cite{Witten2020b} In our experimental observations, bands of type \roto\ achieved Reynolds numbers within the range  $250\le\Re\le300$, owing to their relatively higher terminal velocities. In contrast, bands of types \deve\ and \bscr\ achieved Reynolds numbers in the lower range of $50\le\Re\le250$ due to their comparatively lower terminal velocities. Supplementary videos show the sedimentation of sample bands of each type with various aspect ratios. Notably, bands of type \deve\ consistently exhibited more pronounced tumbling compared to bands of type \bscr\ and \roto. However, the extent of tumbling decreases as the angle $\alpha$ between the cylinders used for constructing the bands increases. Increasing $\alpha$ yields wider inner channels, stabilizing the flow within the folds of the bands. It is important to highlight that triply-twisted bands possess an additional characteristic dimension associated with their vein (or internal opening). Bands of type \deve\ have narrower spaces between folds in comparison to bands of type \roto\ and \bscr, resulting in higher shear rates interior to their undulations. Consequently, inertial instabilities are responsible for the pronounced tumbling motion as  bands of this type sediment. Unlike bands of type \deve, bands of type \roto\ exhibit more stable rotation during sedimentation, with tumbling observed as $\chi$ approaches unity. The petal-like structures of bands of type \roto\ appear to offer greater resistance, contributing to their consistent rotation.

\begin{figure*}[!t]
 \includegraphics[trim=0cm 0cm 0cm 0.cm, clip=true, width=1\textwidth]{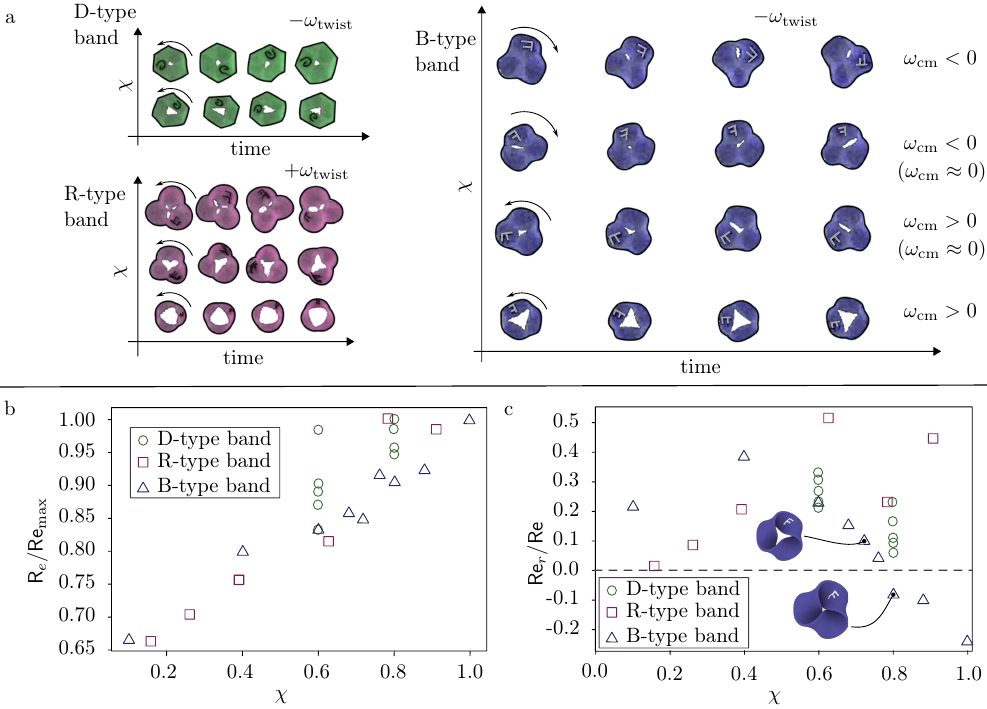}
\caption{
Experimental data on the sedimentation of triply twisted mobius bands. a) Mirrored-bottom images at equally spaced time intervals of developable (\deve), rotoidal (\roto), and binomial scroll (\bscr) type \mob\ bands during sedimentation, and for various aspect ratios $\chi$.  $\chi$ is defined as the ratio $w/w_\text{max}$ where $w$ is the band's width, and $w_\text{max}$ is the maximum band's width before self-intersection. The arrows indicate the direction of rotation.  The spinning direction in mirrored-bottom view coincides with the spinning from the top view. Bands of type \deve\ and \bscr\ constructed with negative twist ($\omega_{\text{twist}}$), whereas bands of type \roto\ positive. b) Ratio of translational Reynold number $\Re$ to maximum translational Reynold number {$\Re_\text{max}$} as a function of $\chi$ for bands of type \deve\ (circles), \roto\ (squares), and \bscr\ (triangles). The {$\Re_\text{max}$} is the Reynold number at $\chi=1$. c) Ratio of rotational Reynold number {$\Re_{r}$} to $\Re$ for bands of type \deve, \roto, and \bscr. The colors of the bands were edited to coincide with the colors in Fig. 1, and the markers on the bands are color enhanced to facilitate their visualization.
}
\label{fig:results3T}
\end{figure*}

We consistently observe rotational motion during the sedimentation of each family of chiral bands. Specifically, each \deve\ band spins in the direction opposite to its twist orientation ($\sgn(\text{spin})=-\sgn (\omega_{\text{twist}})$). Conversely, each \roto\ band spins in the same direction as its twist ($\sgn(\omega_{\text{cm}})=\sgn(\omega_{\text{twist}})$). To further illustrate this phenomenon, we present results in Fig. \ref{fig:results3T}a. In the case of bands of type \deve, those with negative twists undergo a counter-clockwise spinning, while bands of type \roto, which have a positive twist, exhibit positive rotational direction. Bands of type \bscr\ exhibit two different spinning directions, or a \textit{chiral-response switch}, as their aspect ratio varies. Snapshots of \bscr\ bands with negative twists at $\Ar\sim 95400$ and representative choices of $\chi$ are provided in Fig. \ref{fig:results3T}a (see also Supplementary Videos). This \bscr\ chiral-response transition from negative to positive spinning occurs within a narrow window of $\chi$ values, and implies the existence of a critical value $\chi_{\text{crit}}\approx0.8$ where spinning is nearly suppressed ($\omega_{\text{cm}} \approxeq 0$). At $\chi=\chi_{\text{crit}}$, energy dissipation is primarily linked to predominantly vertical translation and possible vortex formation. In Fig. \ref{fig:results3T}b  and c, we present the impact of the parameter $\chi$ on translational ($\Re/\Re_{\text{max}}$) and rotational ($\Rer/\Re$) behaviors for the different types of bands. Here, $\Re_{\text{max}}$ corresponds to $\chi=1$. Notably, \roto\ bands show a nearly monotonic increase in $\Re/\Re_{\text{max}}$, while \bscr\ bands exhibit transitions between $0.7<\chi<0.9$ that coincide with the chiral-response switch. Overall, \bscr\ bands behave like \roto\ bands for $\chi\geq\chi_{\text{crit}}$, while resembling \deve\ bands for $\chi\leq\chi_{\text{crit}}$.

\begin{figure}[!tbhp]
\centering
\includegraphics[trim=0cm 0cm 0cm 0cm, clip=true]{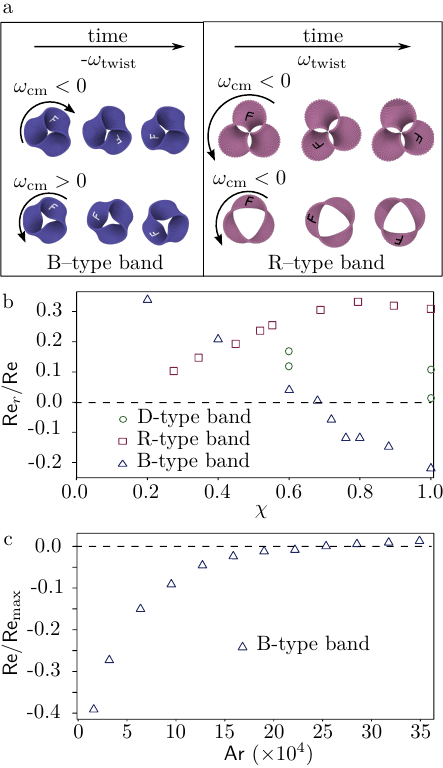}\caption{
Simulation data on the sedimentation of triply twisted mobius bands. a)  Top-view images of bands of type s \bscr\ and \roto\ during sedimentation at equally spaced time intervals, and for two representative values of the ratio $\chi=w/w_\text{max}$, where $w$ represents the width of a band and $w_\text{max}$ is the maximum value of $w$ that can be achieved without self-intersection. b) Ratio of the rotational Reynold number $\Rer$ to the translational Reynold number $\Re$ as a function of the aspect ratio $\chi$. c)  $\Rer\mskip-1mu/\Re$ as a function of the Archimedes number, $\Ar$, which is measures the importance of gravitational forces relative to viscous forces.}
\label{fig:regEffect}
\end{figure}

To simulate the sedimentation behavior of the bands, we use dissipative particle dynamics (DPD).\cite{Groot1997,Moreno2015} A detailed description of the DPD methodology is provided in the Methods section. These simulation studies validate the presence of a chiral-transition regime in bands of type \bscr (see Supplementary Videos), in agreement with our experimental findings. The simulation results demonstrate axial alignment and chirality in bands of this type, as illustrated in Fig. \ref{fig:regEffect}a. However, it is noteworthy that the critical aspect ratio at which the chiral transition occurs shifts to lower values of $\chi$ in simulations compared to experiments, transitioning at $\approx0.68$ as opposed to $0.78$ in experiments (see Fig. \ref{fig:regEffect}b). It is important to acknowledge that certain effects observed in the experiments, such as tumbling, as well as the discrepancy in Reynolds numbers between bands of type \roto\ and \bscr, are not fully replicated in the computational simulations. These discrepancies arise due to inherent limitations of DPD in explicitly enforcing no-slip boundary conditions and in accounting for effects associated with Schmidt number variations. Although the numerical model has limitations in capturing effects such as tumbling and the detailed variations in Reynolds numbers between \roto\ and \bscr\ bands, it still manages to replicate the observed chiral response and its transition seen in experiments. This underscores the dominant influence of shape-related characteristics on sedimentation dynamics, even in the presence of substantial inertial effects. 

\begin{figure*}[!t]
\includegraphics[trim=0cm 0cm 0cm 0cm, clip=true, width=1\textwidth]{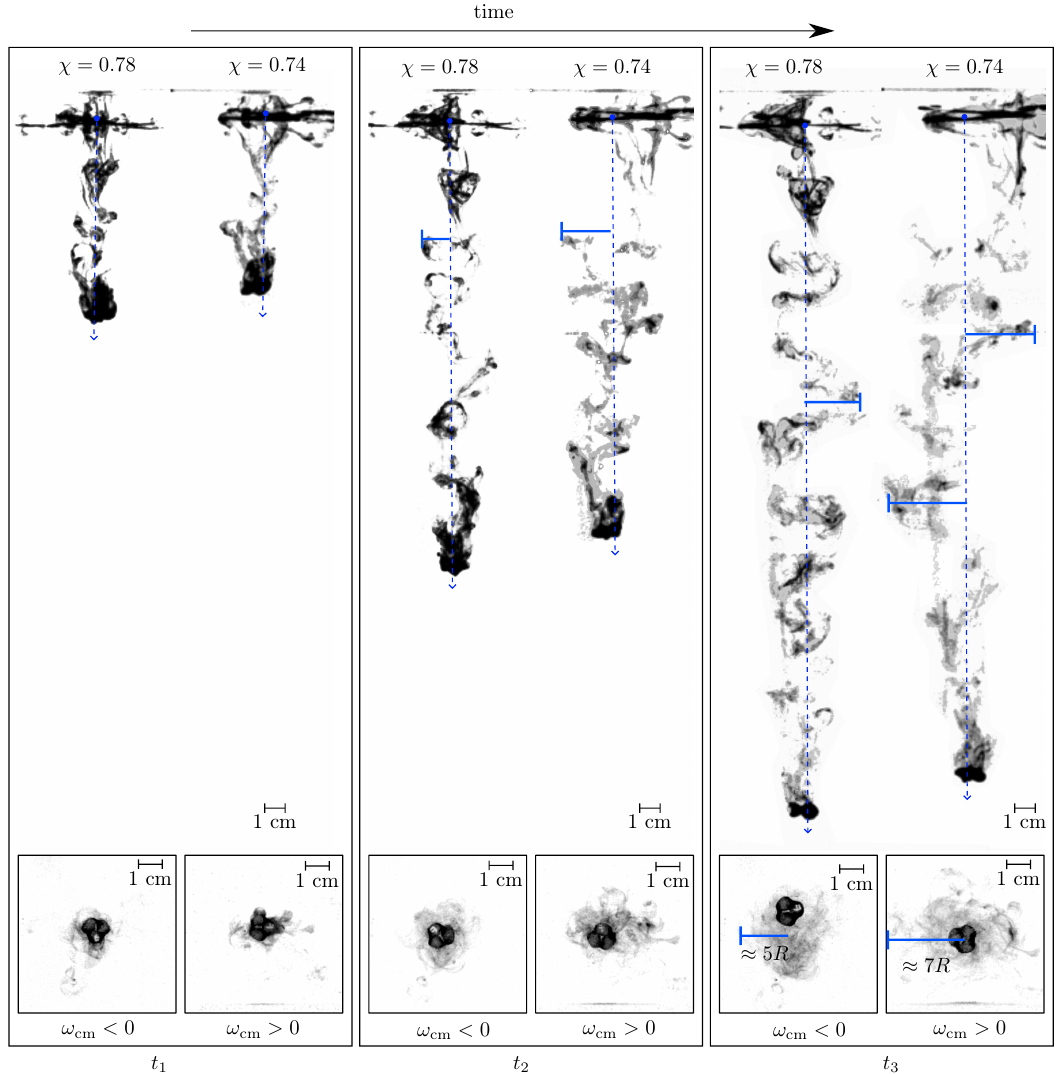}
\caption{
Sedimentation of fluorescein-tinted triply twisted Mobius bands in water. Side view of the sedimentation of bands of type \bscr\ with aspect ratio $\chi=0.78$ and $\chi=0.74$, which exhibit negative and positive angular speed ($\omega_\text{cm}$), respectively, and for increasing time instants $t_1$, $t_2$, and $t_3$. $\chi$ is defined as $w/w_\text{max}$ where $w$ is the band width and $w_\text{max}$ is the maximum band width before self-intersection. Inside images correspond to mirrored-bottom view of the same \mob\ bands at time $t_1$, $t_2$, and $t_3$. The wake created by the bands spand over a distance several times larger than the projected radius $R$. Blue-solid lines indicate the maximum size of the wake. The original images were taken with a black background to enhance the contrast with the green fluorescein. Postprocessing of the images was necessary to obtain a white background with grey scale wake. 
}
\label{fig:results3TBlue}
 \end{figure*}

For our considered range $50\le\Re\le400$ of Reynolds numbers, it has been previously observed that instabilities and wake formation readily occur even for simple objects like spheres.\cite{Johnson1999} To enhance our understanding of the impact of the chiral switch exhibited by bands of type \bscr\ on the suspending fluid, a series of experiments was conducted by initially coating the bands with a fluorescein-salt solution, allowing for the visualization of the fluid flow induced by the sedimenting bands. In Fig. \ref{fig:results3TBlue}, we present the lateral and bottom views of two sedimenting bands of type \bscr\, with different aspect ratio, $\chi<\chi_{\text{crit}}$ and $\chi>\chi_{\text{crit}}$ (see also Supplementary videos). For clarity, the lateral images have been filtered and color-enhanced. Notably, the wake generated by the narrower band exhibits a distinctly different shape. The relative size of the wake is highlighted with blue-solid lines in Fig. \ref{fig:results3TBlue}. Specifically, the wake width produced by the band at $\chi=0.78$ is somewhat more compact ($\approx5R$) in comparison to that generated at $\chi=0.74$ ($\approx 7R$). This observation indicates that at $\chi=0.78$, the band causes less disruption to the fluid, whereas a modest change in the aspect ratio to $\chi=0.74$ results in significant energy dissipation and vorticity generation in the wake. Overall, even a minor change in the aspect ratio of the band seems sufficient to induce modifications in the fluid-induced drag, which, combined with inertial effects, can drive changes in the chiral response of a sedimenting band of type \bscr.


%

\begin{figure*}[!tbhp]
\includegraphics[trim=0cm 0cm 0cm 0cm, clip=true, width=1\textwidth]{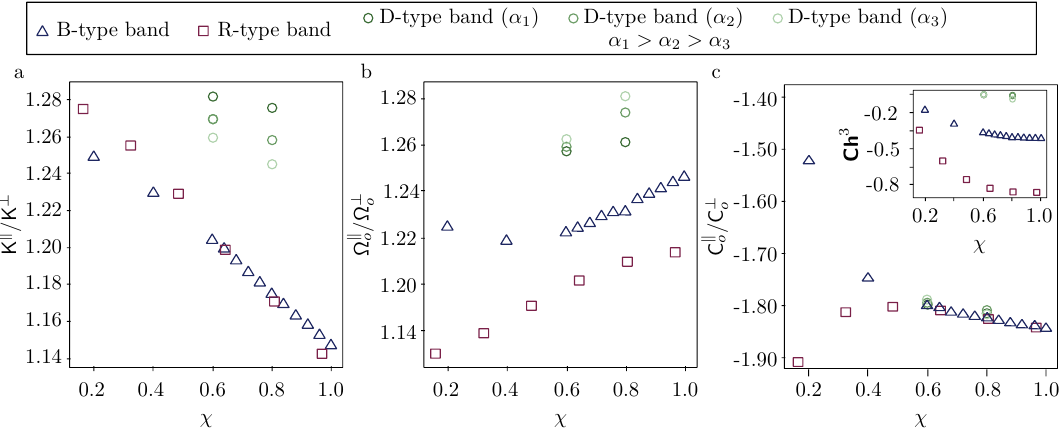}
\caption{
Variation of the components in the resistance tensor, $\tresi$, with respect to the aspect ratio $\chi$ of the bands. Ratio of: a) translational tensor components parallel $\ttranc^{\parallel}$ to perpendicular $\ttranc_{\perp}$, (b) rotational tensor components parallel $\trotc^{\parallel}$ to perpendicular $\trotc^{\perp}$ and c) coupling tensor components parallel $\tcouc^{\parallel}$ to perpendicular $\tcouc^{\perp}$ as a function of $\chi$.  The components are parallel or perpendicular to the gravitational force.
}
	\label{fig:resistance}
\end{figure*}

To explore whether variations in the hydrodynamic drag can serve as an indicator of the chiral switch observed for bands of type \bscr, we employed the rigid multiblob methodology\cite{Usabiaga2016} to calculate the components of $\tresi$ for all three types of bands with various aspect ratios at $\Re\approx0$. Without loss of generality, we present results for bands with negative chirality. For these computations, we take the centroid of the band to be located at the origin, $o$, of an orthonormal basis, $\{\bfe_x,\bfe_y,\bfe_z\}$,  chosen such that $\bfe_z$ is parallel to the vector of the gravitational force (and the axis of rotation). In Fig.~\ref{fig:resistance}a-c, we present the variation in the components of the translational $\ttran$, rotational $\trot$, and coupling $\tcou$ dyads. To highlight disparities in the computed values of the drag along different directions axis, we used the ratio of the components parallel ($\parallel$) and perpendicular ($\perp$) to $\bfe_z$. For instance, for the translational tensor, $\ttranc^{\parallel} = \bfe_z\cdot\ttran\cdot\bfe_z$, whereas $\ttranc^{\perp} = \bfe_x\cdot\ttran\cdot\bfe_x$. Notably, the ratio $\ttranc^{\parallel}/\ttranc^{\perp}$ exhibits a monotonic decrease for bands of type \bscr\ and \roto, while the rotational ratio $\trotc^{\parallel}/\trotc^{\perp}$ shows a moderate variation with  mean values of approximately 1.23 for bands of type \bscr\ and at approximately 1.26 for bands of type \deve. Overall, for large values of $\chi$, bands of type \bscr\ and \roto\ exhibit similar hydrodynamic resistance.  Fig.~\ref{fig:resistance}c provides deeper insight into the behavior of twisted bands, the opposite sign in the components that characterize the coupling resistance ($\tcouc^{\parallel}/\tcouc^{\perp} < 0$) indicate that the chirality of the object depends on the direction from which it is observed. For chiral bodies, the mobility matrix $\tmob = \tresi^{-1}$ characterizes the direction in which that body rotates through the \textit{chirality matrix} $\tch = \tmobc/(\chi\tmobr)$,\cite{Morozov2017, Mirzae2018} where $\tmobc$ and $\tmobr$ are the coupling and rotation dyads of the mobility tensor, respectively. Granted that $\tch$ is expressed relative to a basis $\{\bfe_1,\bfe_2,\bfe_3\}$ defined by the eigenvectors of $\tmobr$, with corresponding eigenvalues enumerated in ascending order, $\tmobrc^1\leq\tmobrc^2\leq\tmobrc^3$, the diagonal elements of $\tch$ are expressed as $\tch^{i} = \tmobc^{ii} / (\chi \tmobr^{i})$, $i=1,2,3$. The inset plot in Fig.~\ref{fig:resistance}c compiles the variation of $\tch^3$ with $\chi$ for each type of band. In general, $\tch$ is monotonically decreasing with the aspect ratio and does not evidence the characteristic transition observed experimentally.  Among the three types of bands, the eigenvalues of $\tch$ have the largest magnitudes for the \roto\ type and the smallest magnitudes for bands of the \deve\ type. For bands of the \bscr\ type, the eigenvalues of $\tch$ fall in between, suggesting a hydrodynamic response that is likely a transitional state between the pronounced characteristics of \roto\ and the subdued features of \deve\ bands. Despite the insights provided by $\tresi$ at low Reynolds numbers, accurately discerning the observed chiral response for bands of the \bscr\ type poses a formidable challenge. The transition between the two distinct spinning directions exhibited by bands of type \bscr\ is  influenced by multiple factors, such as the balances of forces and moments, the alignment of the band, and the onset of inertial instabilities. These complexities highlight the need for further investigations and a comprehensive understanding of the underlying dynamics. 
 

As the transition in chiral response for bands of type \bscr\ is influenced not only by the band's characteristic size $\chi$ but also by alterations in the flow regime, we delve deeper into this phenomenon through additional numerical DPD investigations. For this purpose, we focus on bands of type \bscr\ near the critical aspect ratio, $\chi=0.68$ (prior to the chiral switch in our DPD model), and manipulate the interplay between gravitational and viscous effects by varying the Archimedes number ($\Ar$). In our simulations, this effect is achieved by modifying the external force applied to the band. Fig. \ref{fig:regEffect}c showcases the spinning direction of a \tpt\ \mob\ \bscr\ with $\chi = 0.68$ under varying $\Ar$ numbers. Notably, we observe that the transition in the chiral response of bands of type \bscr\ is a universal phenomenon, consistently present across all the bands. However, the specific point of this transition varies depending on the regime of flow and the characteristic size $\chi$ of the bands.

\section*{Conclusions and Outlook}
~
Our study sheds light on a striking revelation: despite sharing identical topology and intrinsic chirality, triply twisted \mob\ bands with three-fold rotational symmetry can exhibit divergent chiral responses attributed to their unique curvatures. Specifically, our exploration of bands of type \bsc\ uncovers a nuanced interplay of torques and forces that evolve with changing aspect ratios, modulated by the influence of inertial effects. This intricate relationship results in a distinctive transition in the spinning direction of the band. --- a phenomenon absent in other \tpt\ bands. Computational simulations robustly confirm these experimental insights, further elucidating the intricate interdependence between gravitational and viscous forces and the critical aspect ratio dictating the transition.

Beyond the immediate scope of our investigation, our findings bear broader implications. They signal promising prospects in designing responsive materials and microswimmers without self-propulsion, harnessing the potential for hydrodynamically-driven chiral-response switching. This not only points to the creation of deployable objects with dynamic rotational responses but also sparks inquiries into analogous entities with akin attributes and the intriguing concept of hydrodynamically-responsive chirality in objects featuring non-zero coupling tensors. Furthermore, our study underlines the limitations of solely relying on intrinsic chirality to characterize chiral particles when inertial instabilities come into play, thereby prompting further explorations into the intricate tapestry of emergent hydrodynamic interactions.

In closing, our study contributes novel insights to the captivating domain of chiral hydrodynamics, blending fundamental understanding with tangible applications. By deciphering the hydrodynamic intricacies of chiral objects, we can pave the way for innovative technologies and devices endowed with unique attributes. Concurrently, this venture deepens our understanding of the complex behaviors of biological molecules where chirality plays a pivotal role. As the scientific narrative of chiral hydrodynamics continues to evolve, our findings are poised to stimulate further discoveries, facilitate groundbreaking applications, and enrich our grasp of intricate fluid dynamics.

\section*{Methods}

\subsection*{Experimental sedimentation}

To conduct our sedimentation studies, we utilized polystyrene bands with a density of 1.17~g/cm$^3$ in water at room temperature. These bands were 3D-printed using the OBJET 500 from STRATASYS company, and were placed in a container with dimensions of 0.20 x 0.20 x 0.60~m$^3$. To identify the direction of the spin, the bands were marked with the letter F or G. We used two cameras, an Ultrahigh-Speed Camera (v2512) and a High-speed camera (Phantom Miro LC-120), to record the falling of the objects at a rate of 100 frames per second. The cameras were equipped with Nikon AFS 60mm f/2.8G and Sigma 24-70mm f2.8 lenses and were positioned 2 meters away from the sedimentation container. The cameras were aligned to capture 0.4m of the vertical trajectory, while three different markers were placed in the sedimentation container at 10~cm, 25~cm, and 40~cm from the bottom to facilitate trajectory analysis. To record the rotation of the bands, we placed a tilted mirror at the bottom of the container with an angle of $45^o$. The marks (F or G) in sedimenting bands pointed in the sedimentation direction for observation through the mirror, and their spinning direction coincides with top-view observation. The vertical distortion from the center of the field of view up to the edges was $\pm 1.5$~cm, while the horizontal distortion was $\leq 0.4$~cm. The 3D-printed bands were constructed with a positive twist, resulting in a positive spin of their intrinsic chirality. The width of the band was varied in different experiments to investigate the effect of the aspect ratio, while their projected radius and thickness were kept constant at $10$~mm and $0.8$~mm, respectively. The Reynolds numbers evaluated ranged from $50$ to $350$. To extract the position of the bands from the recorded images, we used the distribution of the pixels to track the position of the center of mass of the band in each camera and constructed the three-dimensional trajectories.

\subsubsection*{Image analysis}

To extract the linear and angular velocity of the bands we analysed the images obtained every 60 frames per second. Each image was divided into two regions to recover the lateral and bottom view reflected in the mirror. The image segmentation was then conducted in each of the projections to retrieve the band's pixel count and distribution. Trajectories were reconstructed from the position of the mean distribution of the pixels at each frame. We used the letter marker on the bands to estimate their angular velocity around the vertical axis of a reference frame with the origin at the center of mass of the band.  

\subsection*{Dissipative particle dynamics sedimentation simulations}

We modeled the sedimentation of the bands using the dissipative particle dynamics method (DPD)\cite{Groot1997}. DPD is a particle-based method that consistently captures hydrodynamic effects, and has been widely used to study soft matter in different fields\cite{Moreno2013,Moreno2020}. In DPD, we discretized the \tpt\ bands using a single layer of $N_p$ soft beads rigidly connected, whereas the fluid is discretized using individually interacting particles. An sketch of the discretized version of the different type of bands is presented in SI.Fig. 4. See also SI for the point coordinates of the discretized bands. During the simulations, the force between the fluid and the beads of the bands was computed, and the position and velocity of the center of mass of the band were updated due to the net force acting on it. The position of the $N_p$ beads of the bands was updated following the rigid body dynamics. To conduct these simulations we used the DPD distribution implemented in LAMMPS\cite{Plimpton1995}, as well as the rigid-body motion setting available there. For a detailed description of the DPD method the reader is referred to SI. The visualization of the DPD simulation results was done using the software OVITO\cite{Stukowski2010}.

Sedimentation simulations were conducted in elongated boxes with sizes $L_x=12R$, $L_y=12R$, and $L_z=180R$. All the simulations used periodic boundary conditions on the $x$ and $y$ axis, whereas reflecting walls are used on the $z$-axis. The size $L_x$ and $L_y$ were defined to avoid the effects of the periodic boundary condition. The center of mass $X_{O}$ of bands was initially located at the top of the box at $[0.5L_x, 0.5L_y,0.9L_z]$. The bands were subjected to an external force $f_{\text{g}}$ on the $-z$ direction.  Due to the non-periodicity in $z$, the results of the simulations were analyzed using trajectories in the range $0.8L_z>X_{O}>0.1L_z$, to neglect wall effects.

The interaction parameter of the DPD conservative potential was set as $a_{\textit{bb}} = a_{\textit{ff}} = 25.0$, where $\textit{bb}$ and $\textit{ff}$ correspond to interactions between band-band and fluid-fluid beads, respectively. The interaction parameter between band and fluid beads ($\textit{bf}$) was set as $a_{\textit{bf}} = 65.0$ to ensured no-penetration boundary condition at the surface of the bands. Additionally, the distance between the beads forming the bands was $0.75r_c<r_{\textit{bb}}<r_c$, where $r_c$ is the DPD cutoff radius. These values ensure the control of the density fluctuations across the box and satisfactorily avoid the passing of fluid particles across the band's domains. Additionally, to minimize resolution effects due to the relative size between the band and the discrete size of the fluid, all the bands were constructed such that the width $w$ of the band has to be larger than $5r_c$, and the band radius $R>10r_c$. Non-slip boundary conditions were not explicitly accounted for using the standard DPD methodology, therefore, the simulations cannot be considered strictly no-slip or slip. Simulations were conducted at $k_BT=1$, $m=1$, $r_c=1$, $\rho_n=3$. At the fluid conditions used in our simulations, the viscosity of DPD can be approximated as $0.866$ in reduced units.\cite{Boromand2015}

\subsection*{Resistance tensor calulation}

We used the rigid multiblob (RMB) methodology\cite{Usabiaga2016} to estimate numerically the the resistance tensor $\tresi$ for the different bands at zero Reynolds number. RMB has been successfully applied to study the hydrodynamic behavior of rigid objects with complex morphologies\cite{Moreno2022,Moreno-Chaparro2023}. In the RMB approach, only the set of $N_p$ beads was considered, whereas hydrodynamic interactions with the solvent were accounted for analytically. To compute $\tresi$, we used the discretization adopted for the DPD simulations. However, the beads were treated as rigid bodies (instead of the soft DPD beads), with a hydrodynamic radius $r_{\text{RMB}}=r_{\textit{bb}}/2$. The hydrodynamic interactions between beads were represented using the generalized Rotne--Prager--Yamakawa tensor for an unbounded region.\cite{Rotne1969, Usabiaga2016} For further details concerning the implementation of the RMB, the reader is referred to \cite{Usabiaga2016}.

\textbf{Acknowledgments}
This work was supported by the Okinawa Institute of Science and Technology Graduate University with subsidy funding from the Cabinet Office, Government of Japan. N.M acknowledges the support from the European Union's Horizon 2020 under the Marie Sklodowska-Curie Individual Fellowships grant 101021893, with acronym ViBRheo.

\section*{Declarations}

\subsection*{Competing interests}

The authors declare no competing interests.

\subsection*{Authors contribution}

N.M.~and E.F.~conceived the study. The first draft of the manuscript was written by N.M., and all authors contributed to the final version of the manuscript. N.M~and D.V~conducted the experiments. N.M.~conducted the simulations and data processing. All the authors discussed and analyzed the results. All authors approved the final version of the manuscript.

\subsection*{Availability of data and materials}
Data supporting the findings of this paper are available from the corresponding authors upon request.

\bibliography{thebibliography}

\end{document}